\documentclass[11pt,a4paper]{article}
\usepackage{lipsum}
\usepackage{authblk}
\usepackage{amsmath,a4}  
\usepackage[utf8]{inputenc}
\usepackage{graphicx,enumerate,color,amsfonts,amsthm,amsmath,amssymb,mathrsfs,marginnote,dsfont,textcomp,tikz,setspace,marginnote,subfigure,pdflscape,multirow,booktabs,array,float,enumitem,scrtime,currfile,titleps,hyphenat,relsize,psfrag}
\usepackage[colorlinks,
    linkcolor={red!60!black},
    citecolor={black},
    urlcolor={blue!80!black}]{hyperref}
\usepackage{tikz,pgf,pgffor,pgfplots}
\tikzset{>=latex}
\usetikzlibrary{calc,patterns,decorations,intersections,decorations.pathmorphing,fit,backgrounds,arrows,shapes,automata,petri,positioning}
\usepgfmodule{shapes,plot}

\theoremstyle{definition}

\theoremstyle{remark}

\makeatletter
\newcommand{\setword}[2]{%
  \phantomsection
  #1\def\@currentlabel{\unexpanded{#1}}\label{#2}%
}
\makeatother

\newcommand{\f}{\textnormal{\textbf{f}}}
\newcommand{\g}{\textnormal{\textbf{g}}}

\newcommand{\e}{\mathrm{e}}
\newcommand{\R}{\mathbb{R}}
\newcommand{\RM}{\mathcal{R}}
\newcommand{\OM}{\mathcal{O}}
\newcommand{\DM}{\mathcal{D}}
\newcommand{\VM}{\mathcal{V}}

\allowdisplaybreaks
\usepackage[top=2cm, bottom=2cm, left=2cm, right=2cm]{geometry}
\usepackage{fancyhdr}
\newpagestyle{main}{
\setheadrule{.4pt}
\sethead{}
        {}
        {\sectiontitle}
}

\renewenvironment{abstract}{%
\hfill\begin{minipage}{0.95\textwidth}
\rule{\textwidth}{1pt}}
{\par\noindent\rule{\textwidth}{1pt}\end{minipage}}
\makeatletter
\renewcommand\@maketitle{%
\hfill
\begin{minipage}{0.95\textwidth}
\vskip 2em
\let\footnote\thanks 
{\Large \bf \@title \par }
\vskip 1.5em
{\large \@author \par}
\end{minipage}
\vskip 1em \par
}
\makeatother

\begin{document}
%
\title{\nohyphens{An epidemic model highlighting humane social awareness and vector--host lifespan ratio variation$^*$}}
\author[1]{Karunia Putra Wijaya}
\author[2,1]{Joseph P\'aez Ch\'avez}
\author[3,$\ast$]{Dipo Aldila}

\affil[1]{\small\emph{Mathematical Institute, University of Koblenz, 56070 Koblenz, Germany}}
\affil[2]{\small\emph{Center for Applied Dynamical Systems and Computational Methods (CADSCOM), Faculty
of Natural Sciences and Mathematics, Escuela Superior Polit\'ecnica
del Litoral, P.O. Box 09-01-5863, Guayaquil, Ecuador}}
\affil[3]{\small\emph{Department of Mathematics, University of Indonesia, 16424 Depok, Indonesia}}

\affil[$\ast$]{Corresponding author. Email: \href{mailto:aldiladipo@sci.ui.ac.id}{aldiladipo@sci.ui.ac.id}}
\maketitle
\begin{abstract}
Many vector-borne disease epidemic models neglect the fact that in
modern human civilization, social awareness as well as self-defence
system are overwhelming against advanced propagation of the disease.
News are becoming more effortlessly accessible through social media
and mobile apps, while apparatuses for disease prevention are
inclined to be more abundant and affordable. Here we study a simple
host--vector model in which media-triggered social awareness and
seasonality in vector breeding are taken into account. There appears
a certain threshold indicating the alarming outbreak; the number of
infective human individuals above which shall actuate the
self-defence system for the susceptible subpopulation. A model where the
infection rate revolves in the likelihood of poverty, reluctancy,
tiresomeness, perceiving the disease as being easily curable, absence of medical access, and overwhelming hungrier
vectors is proposed. Further discoveries are made from undertaking
disparate time scales between human and vector population dynamics.
The resulting slow--fast system discloses notable dynamics in which
solution trajectories confine to the slow manifold and critical
manifold, before finally ending up at equilibria. How coinciding the
slow manifold with the critical manifold enhances periodic forcing
is also studied. The finding on hysteresis loops gives insights of
how defining alarming outbreak critically perturbs the basic
reproductive number, which later helps keep the incidence cycle on
small magnitudes.
~
\end{abstract}

\section{Introduction}
\label{intro} World population has witnessed social and monetary
misfortunes from the spreading of vector-borne diseases since
subsequent centuries \cite{gubler1991,gubler1998,Gub2009}. Many intervention strategies have been
researched and implemented to fight against the diseases, most of
which are based on suppressing vector population and shielding
humans from contacts with vectors \cite{Vincent2016,Kamareddine2012}. Despite learnable
seasonality of the prominent meteorological factors, therefore of
the vector population, the disease-related incidences continue to
remain cyclical \cite{WAS2019}. Questionable here are thus, the sureness
and regularity in implementing such intervention strategies.

Media reports have played a significant role in influencing
individuals' states of mind and practices during epidemics \cite{Salathe2012, Zhou2019},
keeping them aware of surrounding infection threats. Such
computerized information nowadays is openly accessible from numerous
sources including direct news from broad communications (e.g.\
radios, televisions, newspapers, booklets) and catchphrases/hashtags
from social media (e.g.\ Facebook, Twitter, Instagram). The latter
can especially gauge further detailed information including
geospatial labels into a certain sphere only under short typing. The
main task of the media in the context of anti-disease campaign is to
scatter incidence data and regularly flag up related themes
including possible causes, symptoms, worsening effects, data
forecasts, clinical accessibility, prevention strategies, and
emergent solutions \cite{Collinson2014,Collinson2015}. It is a moderately modest method for
reporting centralized information regarding neighbours' wellbeing
and can certainly return much broader influence,
notwithstanding that outbreak data can be hardly accessible through
personal approaches. In response, individuals educated by
media can play safe ranging from injecting vaccines, smearing
repellent fluids, wearing defensive clothing, to staying away from
social contacts with infected humans and from endemic regions
\cite{Collinson2015}. Educated infective humans may likewise take measures to ban
themselves from being exposed to others to diminish infectivity.

Recently, a number of modeling studies have been done to evaluate
the impacts of media reports on the change of individual conducts
against the spread of infectious diseases. Except the statistical distribution-matching \cite{YHB2010}, classical linear regression \cite{RR2007} and game-theoretic approach
\cite{Che2009}, the existing mathematical models in this context engineer
differential equations -- typically SIR-type models -- as to govern
incidence pattern. The latter mostly fall into two ideas. The first
idea highlights media as to give feedback to a system for the
infectivity lessens as the number of infective individuals gets
larger. Placement of the corresponding measures depends on the types
of actions taken against the disease spread. In case of
vaccination-like preventive actions, the feedback serves as a rate
in taking up susceptible hosts \cite{SM2014}. In case of repellence
against contacts, it usually serves as modification of the infection
or contact rate. To this later case, the infection rate is likely to
be a decreasing function of the infective (and exposed)
subpopulation, which can be either a rational function \cite{LC2008,Li2009,Zhao2018}, an
exponential function \cite{CSZ2007,Liu2007,Zhou2019} or a rather generalized version \cite{CTZ2008}. The second idea includes the
introduction of ``aware'' subpopulation from the original
susceptible, infected, and recovered subpopulation \cite{Agaba2017,Basir2018,Misra2011}. The rate
at which an ``unaware'' individual becomes ``aware'' can thus be
modeled as an (increasing) function of the infective host
subpopulation \cite{SM2014} and/or a certain measure for the intensiveness
of media reports \cite{MSS2011,Misra2011,GRS2015,Agaba2017}. Another view also sees a two-way
relationship, as such intensiveness increases along with an
increasing number of incidences \cite{MSS2011,Agaba2017}.

In this paper, we present a model that follows the first idea. The
governing equations are, possibly the simplest SISUV model with
constant host population and saturating vector population. A novelty
here is the introduction of an alarming incidence level $j^{\ast}$,
below which medical departments can never transfer information to
media holders for either time, interest, or financial restrictions.
We further develop two models for the infection rate. The first
model portrays a non-increasing infection rate, which is based on
the situation where the hosts keep up the pace in taking up
preventive measures along with ever-streaming media reports.
Notwithstanding different treatment in the model, previous
investigation \cite{LC2008} equivalently indicates the supercritical-type
of bifurcation of the model system. The second model considers the
scenario where the alarming outbreak $j^{\ast}$ is defined as the
maximum number of patients the available hospitals in the observed
region can accommodate. The case that the disease is endemic in
``developing'' regions also sets additional factors why human's
exposure to infection can get higher with the incidence level. The
infection rate accordingly decreases due to media reports, yet it
revolves as the incidence level gets higher due to poverty,
reluctance in taking up preventive measures, tiresomeness, perceiving the disease as being easily curable, absence
of medical access, and presence of hungrier vectors.

For more realistic touching, we include a seasonal forcing in the
vector population due to meteorological factors. Periodical climatic
patterns have been argued to be one of the most influential
conditions that catalyze the infection processes \cite{Babin2003,Bartley2002}. This stems
from the observation that disease vectors essentially look for the
most favorable ambient temperature, humidity, wind speed and water
precipitation surrounding their life cycle \cite{Bicout2015,Cheong2013}. In some tropical
and subtropical regions, for example Jakarta, Indonesia \cite{WAS2019},
Taiwan \cite{WGL2007} and Sisaket, Thailand \cite{WJJ2013}, meteorological factors
might be too random under a small time scale, but they often exhibit
apparent long-term trends with certain periodicities. At this point,
the behavior of meteorological factors, and therefore that of
incidence levels, become more ``understandable''. This fact gives us
useful information for more accurate forecasts and the
implementation of disease controls.

However, due to a natural discrepancy on the lifespans of vector and
host, such model might portray significantly different solution
trajectories under the variation of the lifespan ratio $\epsilon$.
Added with another assumption on the infection rates, the ratio
variation gives birth to a singularly perturbed system. This allows
the field of the vector dynamics to entirely be controlled by
$\epsilon$. Two traditional results are underlying: critical
manifold, the surface representing the equilibrium of the system
under the assumption that the vector lifespan is infinitesimal
($\epsilon=0$), and slow manifold, a (locally attractive) surface as
perturbation of the critical manifold in case of full stability for
$\epsilon>0$. Therefore, as the reminder we go towards answering the
following questions: What is the stability status of the critical
manifold? How can solution trajectories confine to the slow
manifold, before approaching the critical manifold and ultimately a
stable equilibrium? How is the slow manifold approximated? How many
are and how are the stability statuses of the endemic equilibria?
How do the periodic solutions behave with respect to the basic
reproductive number and the alarming incidence level? How to see if
the periodic solutions expand with the amplitude of the seasonal
forcing? What happens to the periodic solutions in case
$\epsilon\rightarrow 0$ and $\epsilon\lesssim 1$?

\section{Model derivation}\label{sec2}

Let $N:=S+I$ denote a total host population size on the observed
region, which is assumed to be constant due to a relatively tiny
increment rate on a usual time scale of the vector dynamics. This
population size shares a time-dependent population size of
susceptible hosts $S$ and that of passively infective hosts $I$.
Analogously, $M:=U+V$ denotes a total vector population size,
comprising a population size of susceptible vectors $U$ and that of
actively infective vectors $V$. Our point of departure in the
modeling consists in reducing the following SISUV model
\begin{equation}\label{eq:SISUV}
\begin{split}
S'&= \mu(N-S)-\tilde{\beta} SV+\gamma I,\\
I'&=\tilde{\beta}
SV-(\gamma+\mu)I,\\
U'&=\Lambda -\tilde{\rho} UI-\theta U,\\
V'&=\tilde{\rho} UI-\theta V.
\end{split}
\end{equation}
Here, $\mu$ denotes the host natural mortality rate, assumed to be
the same as the natural natality rate for the sake of the constancy
of $N$. In the vector dynamics, $\Lambda,\theta$ denote the
recruitment rate and natural mortality rate, respectively. The
parameters $\tilde\beta$ and $\tilde{\rho}$ denote the rate of
infection from an infective vector to a susceptible human and that
from an infective human to a susceptible vector, respectively. The
parameter $\gamma$ denotes the recovery rate that exclusively
contains information regarding loss of immunity. As a specific
feature of the model, we highlight the dependency of the vector
reproduction to the seasonally periodic climatic factors. Taking
into account only climatic factors of \emph{commensurable} periods,
i.e.\ those that share rational dependencies, we can model (cf. \cite{WAS2019})
\begin{equation*}\label{seasonality}
\Lambda = \xi + \zeta \cos (2\pi\omega t)
\end{equation*}
where $\xi,\zeta,\omega,\omega^{-1}$ denote an intercept, an
amplitude, a frequency and the corresponding period, respectively.
On the view of the selected climatic factors, $\omega^{-1}$ shall be
associated with the least common multiple of the commensurable
periods. The amplitude $\zeta$ serves as a tuning parameter for the
importance of seasonality. For the sake of well-posedness and
simplicity, we assume that
\begin{equation}\label{eq:zeta}
0\leq \zeta <\xi.
\end{equation}
Accordingly, the total vector population satisfies
\begin{equation*}
M'=\xi + \zeta \cos (2\pi\omega t)-\theta M,\quad M(0)=M_0.
\end{equation*}
The above equation leads to the exact solution
\begin{equation}\label{eq:M}
M=\left(M_0-\bar{M}-\zeta A_c\right)\e^{-\theta t}+\bar{M}+\zeta
A_c\cos(2\pi\omega t)+\zeta A_s\sin(2\pi\omega t),
\end{equation}
where
\begin{equation}\label{eq:theta}
\bar{M}:=\frac{\xi}{\theta},\quad
A_c:=\frac{\theta}{\theta^2+4\pi^2\omega^2},\quad
A_s:=\frac{2\pi\omega}{\theta^2+4\pi^2\omega^2}.
\end{equation}
It is clear that the population size converges to a periodic
solution as $t\rightarrow\infty$.

For the sake of scaling, we shall divide the host and vector
dynamics with reference constants. In the host dynamics, $N$ would
be the usual choice. In the vector dynamics, we appoint the average
of $M$ on one full period $[0,\omega^{-1}]$. However, the first term
in $M$ as in \eqref{eq:M} makes the averaging varying depending on
the domain undertaken. Therefore, we restrict $M_0$ as to satisfy
$M_0-\bar{M}-\zeta A_c=0$ so that $M$ becomes periodic and yield
\begin{equation*}\label{assumption}
\omega\int_0^{\omega^{-1}}M(t)\,\text{d}t=\bar{M}\quad\text{and}\quad
\omega\int_0^{\omega^{-1}}V(t)\,\text{d}t\leq \bar{M}.
\end{equation*}
The latter holds due to the fact that the nonnegative orthant is
invariant under the flow of \eqref{eq:SISUV}. If seasonality is
negligible ($\zeta=0$), then one yields an identity $M=M_0=\bar{M}$.
We thus define the following scaling
\begin{equation}\label{eq:h}
\frac{M}{\bar{M}}=1+\zeta h(t),\quad\text{where}\quad
h(t):=\frac{A_c}{\bar{M}}\cos(2\pi\omega
t)+\frac{A_s}{\bar{M}}\sin(2\pi\omega t).
\end{equation}
It is apparent to see that $h$ is $\omega^{-1}$--periodic. Under the
following definitions
\begin{equation*}
s:=\frac SN, \quad j:=\frac IN, \quad u:=\frac U{\bar M}, \quad
v:=\frac V{\bar M},\quad\beta:=\tilde{\beta}\bar{M},\quad
\kappa:=\gamma+\mu,\quad \rho:=\tilde{\rho} N
\end{equation*}
together with the constancy of $N$, the system \eqref{eq:SISUV}
reduces to
\begin{equation}\label{eq:jv}
\begin{split}
j'&=\beta(1-j)v-\kappa j,\\
v'&=\rho(1+\zeta h-v)j-\theta v.
\end{split}
\end{equation}
The time scale $t$ in the model~\eqref{eq:jv} is defined on weekly
basis.

According to Esteva--Vargas \cite{EV1998}, both $\beta$ and $\rho$ convey
entities that lead the contact between host and vector to successful
infection. These include the mosquito biting rate and an effectivity
measure representing how successful a mosquito bite leads to virus
transmission. The former is dependent on hosts' mobility that leads
them to sites where the vector population concentrates and how
exposed their skins are. The latter is what we can assume to be
constant on the population level, even though empirical evidence
shows its dependence on age \cite{JC2010}. We assume that most infected
hosts are hospitalized, meaning that they are kept in isolated,
hygienic rooms where vectors are less likely to present.
Consequently, either more hosts are hospitalized or more infective
vectors are surrounding hospitals cannot change the mode of mosquito
bites to infected hosts. It thus is justifiable to assume that
$\rho$ is constant. As far as $\beta$ is concerned, it is the aim of
the current study to model $\beta$ as a function of the infected
host class $j$, i.e.\ taking into account the social awareness
between susceptible hosts in the observed region.
Sections~\ref{sec:report}--\ref{sec:nonreport} are devoted to the
corresponding discussions. In what follows, however, a general
$\beta=\beta(j)$ apparently affords some preliminary analyses, which
will be used in the subsequent sections.

\section{Analysis under time scale separation}
\subsection{Assumptions leading to time scale separation}\label{sec:asump}
As a first step we assume that $\rho,\theta$ are unobservable.
Suppose that data on infected hosts and infected vectors are given,
where both fluctuate at certain orders of magnitude, around certain
medians $(\bar{j},\bar{v})$. Suppose that $\theta$ is pre-specified.
On the virtue of data assimilation, $\rho$ can be traced. At this
stage, we may assume that the data are not heavily fluctuating,
since then $\rho\slash \theta\approx \bar{v}\slash
(\bar{j}(1-\bar{v}))=\text{constant}$ due to Euler approximation on
$v$--dynamics in \eqref{eq:jv} under negligible seasonality
($\zeta=0$). A large $\theta$ in the model returns significant
natural deaths, implying smaller vector lifespan,
therefore $\rho$ has to be chosen equivalently large to keep the
model solution portraying the data. When $\theta$ is assigned with a
smaller value, or larger vector lifespan, then a smaller
force of infection $\rho$ would be preferable to keep the model
solution at the same order of magnitude as when using a larger
$\theta$. From the system, we deduce that the host and vector
lifetime duration fulfil the condition $\mu^{-1}\gg \theta^{-1}$,
making both dynamics run on disparate time scales. Accordingly,
there exists an \emph{adiabatic parameter} $\epsilon$ satisfying
\begin{equation}\label{eq:epsilon}
0<\epsilon\ll 1
\end{equation}
such that $\theta=\mu\slash \epsilon$. By such definition, the adiabatic parameter $\epsilon$ can also be the \emph{vector--host lifespan ratio}. Since $\rho\slash \theta$ is
constant where both $\rho$ and $\theta$ are variable, there exists a
parameter $\rho_{\epsilon}$ such that $\rho\slash
\theta=\rho_{\epsilon}\slash \mu=\rho_{\epsilon}\slash (\epsilon
\theta)$, implying $\rho_{\epsilon}=\epsilon \rho$. It is assumed
that $\rho_{\epsilon},\mu$ be specified beforehand, while
$\rho,\theta$ adjust accordingly based on the variation of
$\epsilon$. Here we present numerical values of the parameters
involved in the model, except where $\epsilon$ and $\zeta$ vary,
satisfying \eqref{eq:epsilon} and \eqref{eq:zeta} respectively.
\begin{table}[htbp!]
\centering \resizebox{\textwidth}{!}{
\begin{tabular}{ccccccccccc}\toprule
$\mu^{-1}$ & $\gamma^{-1}$ & $\rho_{\epsilon}$ & $\theta$ & $\rho$ &
$\xi$ & $\omega^{-1}$ & $\bar{M}$ & $A_c$ & $A_s$ & $M_0$\\ \hline
$[\text{w}]$ & $[\text{w}]$ & $[\text{w}^{-1}]$ & $[\text{w}^{-1}]$
& $[\text{w}^{-1}]$ & $[\text{mos.}\times\text{w}^{-1}]$ &
$[\text{w}]$ & $[\text{mos.}]$ & $[\text{w}]$& $[\text{w}]$ &
$[\text{mos.}]$\\ \hline
 $75\times 48$ & $24$ & $1.8\times 10^{-5}$ & $\mu\slash\epsilon$ & $\rho_{\epsilon}\slash\epsilon$ & $10^{4}$ & $52$ & $\xi\slash\theta$ &
 $\frac{\theta}{\theta^2+4\pi^2\omega^2}$ & $\frac{2\pi\omega}{\theta^2+4\pi^2\omega^2}$ & $\bar{M}+\zeta A_c$ \\ \bottomrule
\end{tabular}
} \caption{\label{tab:par}Parameter values and units used in the
model simulations.}
\end{table}

The appearance of the adiabatic parameter $\epsilon$ also rescores the
amplitude of the seasonal forcing in the model \eqref{eq:jv}. Let
$\text{Ampl}[\cdot]$ denotes the maximal amplitude of a functional
argument with periodic behaviour. We get
\begin{equation*}
\text{Ampl}[h]=\sqrt{\frac{A_c^2+A_s^2}{\bar{M}^2}}=\frac{1}{\bar{M}\sqrt{\theta^2+4\pi^2\omega^2}}=\frac{\mu}{\xi\epsilon\sqrt{\theta^2+4\pi^2\omega^2}},
\end{equation*}
where $h$ is as given in \eqref{eq:h}. We would thus like to study
the possible impact of letting $\epsilon$ varying on the periodic
solutions emanating from $\zeta>0$. Further consequence reveals that
\begin{equation*}
\text{Ampl}\left[\frac{M}{\bar{M}}\right]=1+\zeta\cdot\text{Ampl}[h]=1+\left(\frac{\mu}{\xi\sqrt{\theta^2+4\pi^2\omega^2}}\right)\left(\frac{\zeta}{\epsilon}\right)\sim\frac{\zeta}{\epsilon}.
\end{equation*}
We see here that the maximal amplitude of the seasonal forcing
$M\slash \bar{M}$ is equivalent to the amplitude $\zeta$, but
inversely equivalent to $\epsilon$. As things develop, we will see
how the $(\zeta,\epsilon)$-variations lead to distinctive maximal
amplitudes of not only vector trajectories, but also host
trajectories.

\subsection{Slow--fast system in the absence of seasonality}\label{sec:slowfast}
The point of departure in the analysis is to see how the autonomous
system behaves with respect to the adiabatic parameter $\epsilon$.
From the time scale separation, we discover a singularly perturbed
system
\begin{equation}\label{eq:slowfast}
\left.\begin{array}{rl}
j'&\!\!\!\!=J_0(j)+J_1(j)v\\
\epsilon v'&\!\!\!\!=V_0(j)+V_1(j)v
\end{array}\!\!\right\}x'=\f(x)=(\f^j,\f^{v})(x)
\end{equation}
where
\begin{equation*}
J_0(j):=-\kappa j,\quad J_1(j):=\beta(j)\cdot (1-j),\quad
V_0(j):=\rho_{\epsilon}j,\quad V_1(j):=-\rho_{\epsilon}j-\mu,
\end{equation*}
and $j,v$ correspond to the \emph{slow} and \emph{fast dynamics},
respectively. The \emph{critical manifold} of this system is
characterized by the curve $(j,v^{\ast}(j))_{j\in\DM}$ where
$v^{\ast}(j)=-V_0(j)\slash V_1(j)=\rho_{\epsilon}j\slash
(\rho_{\epsilon}j+\mu)$ and $\DM$ is a connected subset of $[0,1]$.
The according slow dynamics should then be governed by
$j'=\f^j(j,v^{\ast}(j))$ in the critical manifold. This vector field
is continuously differentiable with bounded derivative, guaranteeing
the existence and uniqueness of $j$. Since
$\left.\partial_v\f^{v}\right|_{(j,v^{\ast})}=V_1(j)<0$, then the
critical manifold is \emph{normally hyperbolic} \cite{Kue2015} and moreover, asymptotically stable.
We use $v\mapsto \VM(j,v):v\mapsto\frac{1}{2}(v-v^{\ast})^2$ for the
according Lyapunov function. It is clear that $\VM$ has the
nondegenerate minimum at the critical manifold, where
$\VM'=(v-v^{\ast})\f^{v}\slash\epsilon=-(\rho_{\epsilon}j\slash
\epsilon+\mu\slash\epsilon)(v-v^{\ast})^2\leq -2(\mu\slash\epsilon)
\VM$ owing to $j\in \DM$. Due to $0<\epsilon\ll 1$, there exists a
positive constant $L$ where the inequality $\epsilon \VM'\leq -2\mu
\VM+2\mu L\epsilon \sqrt{\VM}$ holds still. Dividing both sides by
$2\sqrt{\VM}$ and solving the differential inequality forward in
time, we obtain the \emph{slaving condition} for the critical
manifold
\begin{equation}\label{eq:slaving}
|v(t)-v^{\ast}(t)|\leq K|v_0-v^{\ast}_0|\e^{-\mu t\slash
\epsilon}+L\epsilon,\quad v^{\ast}_0=\rho_{\epsilon}j_0\slash
(\rho_{\epsilon}j_0+\mu)\text{ and }K>0.
\end{equation}
This shows that the fast dynamics evolve to the critical manifold
with respect to the \emph{slow time scale} $t$. Moreover, at
$t=\OM(\epsilon|\!\log\epsilon|)$, any fast dynamics starting from a
neighbourhood of order $\OM(1)$ of the critical manifold reaches a
neighbourhood of order $\OM(\epsilon)$ of it. This is in agreement
with the standard result from Tikhonov \cite[Theorem~11.1]{Kha2002}. Under the slaving
condition~\eqref{eq:slaving}, $(t,\epsilon)$--approximation to the solution of
\eqref{eq:slowfast} can be calculated using e.g.\
O'Malley--Vasil'eva expansion \cite{Ver2007}. This facilitates an easier
way to approximate the solution by dissevering the calculation into
those with respect to orders of $\epsilon$.

Here we skip the asymptotic expansions and take a step further. We
are instead looking for an intermediate locally attractive manifold,
to which all nearby solution trajectories confine, which ultimately
coincides with the critical manifold as $\epsilon=0$. This
intermediate manifold is what is known as the \emph{slow manifold}.
The fact that the critical manifold is asymptotically stable,
Fenichel \cite{Fen1979,Jon1995,KK2002} shows that it perturbs with
$\epsilon>0$ to the slow manifold. We
reemploy the slow--fast system \eqref{eq:slowfast} and obtain an
approximation of $v$--state in the slow manifold $v=v(j,\epsilon)$
using center manifold analysis around the critical manifold. To do
so, we first keep away $\epsilon$ from appearing in front of the
derivative term by introducing a \emph{fast time scale}
$\tau:=t\slash \epsilon$ to the system. Note that, again, $t$ shall
be first specified and $\tau$ adjusts accordingly. The system is
then equivalent to
\begin{equation}\label{eq:fast}
\begin{split}
j'&= \epsilon\left(J_0(j)+J_1(j)v\right),\\
v'&= J_0(j)+J_1(j)v,\\
\epsilon'&=0.
\end{split}
\end{equation}
Now the apostrophe indicates the time derivative with respect to
$\tau$. The last system admits the concatenated critical manifold
$(j,v^{\ast},0)$ as its non-hyperbolic equilibrium, i.e.\ the
Jacobian has the eigenvalues
$0,\left.\partial_v\f^{v}\right|_{(j,v^{\ast})},0$. Of course,
$v^{\ast}$ is the equilibrium of the $v$--dynamics in
\eqref{eq:fast} and is asymptotically stable due to
$\left.\partial_v\f^{v}\right|_{(j,v^{\ast})}<0$. This asymptotic
stability of the critical manifold could have never been achieved
unless $v^{\ast}$ is asymptotically stable in the center manifold.
Locally, a slow manifold
$(j,v(j,\epsilon),\epsilon)_{(j,\epsilon)\in\DM_{\epsilon}}$, where
$\mathcal{D}_{\epsilon}\subset [0,1]^2$, acts as a center manifold
and the asymptotic stability of the critical manifold makes it
locally attractive. According to Center Manifold Theorem, there
exists a realization of the slow manifold
$v(j,\epsilon)=v^{\ast}+\OM(\lVert (j,\epsilon)\rVert^2)$, which can
be written in the following abstraction
\begin{equation}\label{eq:approxv}
v(j,\epsilon)=\sum_{i\geq 0}\epsilon^i\g_i(j),
\end{equation}
where $\g_0=v^{\ast}$. According to either \eqref{eq:slowfast} or
\eqref{eq:fast}, this ansatz solves the partial differential
equation
$\f^v=\text{d}_{\tau}v=\partial_jv\text{d}_{\tau}j+\partial_{\epsilon}v\text{d}_{\tau}\epsilon=\epsilon\partial_jv\f^j$
due to $\text{d}_{\tau}\epsilon=0$, which is equivalent to
\begin{equation}\label{eq:pde}
V_0+V_1v(j,\epsilon)=\epsilon
\partial_jv(j,\epsilon)\left(J_0+J_1v(j,\epsilon)\right).
\end{equation}
To calculate the slow manifold $v(j,\epsilon)$ numerically, we
require to fix $\epsilon$ and have certain known point(s) as the
initial condition. If $v(j,\epsilon)=v^{\ast}(j)$, then the
left-hand side of \eqref{eq:pde} vanishes and the right-hand side of
it leaves us either $\text{d}_jv^{\ast}=0$, which can never be the
case for arbitrary $j$, or $J_0+J_1v^{\ast}=0$. The latter
supplements with the fact that the slow manifold intersects with the
critical manifold at the equilibria of the slow dynamics.
Unfortunately, the only reliable points for the initial conditions
are the equilibria, but they give indefinitenesses to start the
computation, $\partial_jv=0\slash 0$. Due to this reason we opt to
approximate the slow manifold. The job is done by rearranging
\eqref{eq:pde}, whereby
\begin{equation*}
-V_0\epsilon^0+\mathlarger{\sum_{i\geq
0}}\left(-V_1\g_i\right)\epsilon^i+\left(J_0\text{d}_j\g_i+\frac{J_1}{2}\text{d}_j\sum_{\substack{m,n\geq
0:~m\neq
n,\\m+n=i}}\g_m\g_n\right)\epsilon^{i+1}+\left(\frac{J_1}{2}\text{d}_j\g_i^2\right)\epsilon^{2i+1}=0.
\end{equation*}
At the expense of vanishing all the coefficients of $\epsilon^i$,
one further yields
\begin{equation*}
0=-V_0\delta_{i0}-V_1\g_i+J_0\text{d}_j\g_{i-1}+\frac{J_1}{2}\text{d}_j\sum_{\substack{m,n\geq
0:~m\neq
n,\\m+n=i-1}}\g_m\g_n+\frac{J_1}{2}\text{d}_j\hat{\g}^2,\quad\text{where}\quad
\hat{\g}=\begin{cases}
\g_{\frac{i-1}{2}},&i\text{ odd}\\
0,&i\text{ even}\\
\end{cases}.
\end{equation*}
We have used $\delta_{ij}$ denoting the usual Kronecker delta. This
last equation is a linear equation in $\g_i$ whose solution can be
calculated straightforward and exhibits a recursion relation.
Surely, disclosing more higher orders is possible but laborious. In
the sequel, we will see how solution trajectories of the model
approach the slow manifold, before approaching the critical manifold
and ultimately approaching stable equilibria of the slow
dynamics in the critical manifold. Numerically approximated slow
manifolds in the virtue of the above discussion will also be
displayed alongside.

\subsection{Periodic solutions of the full system}
This section presents a check for the existence of
$\omega^{-1}$--periodic solutions of~\eqref{eq:jv} under the
activation of seasonal forcing $\zeta>0$, which will be used in the
subsequent discussions. The basic idea highlighting the result has
been adopted from \cite{Sid2013}. Let $\hat{x}$ be an existing equilibrium
of the autonomous counterpart $x'=\f(x;\zeta=0)$. Suppose that we
impose an initial condition $x_0\in U_1(\hat{x})$ for some
neighbourhood $U_1(\hat{x})$ and $|\zeta|<\zeta_1$ for some
$\zeta_1$ such that a unique solution $x=x(t;x_0,\zeta)$ exists. Now
we are looking for an existing periodic solution $x(t;x_0,\zeta)$
surrounding the equilibrium $\hat{x}$, i.e.\
$x(t+\omega^{-1};x_0,\zeta)=x(t;x_0,\zeta)$ for all time $t\geq 0$.
This can be rephrased to looking for a suitable $\zeta$--dependent
initial condition $x_0(\zeta)$ that leads to the periodic solution.
A first step to this, an auxiliary function $S(x_0,\zeta):=
x(\omega^{-1};x_0,\zeta)-x_0$ is set up towards finding $x_0(\zeta)$
that zeros $S$ using Implicit Function Theorem around the point
$(\hat{x},0)$.  Owing to regularity of $\beta$, the vector field
$\f$ becomes continuously differentiable in time $t\in\R_+$, state
$x\in [0,1]\times\R_+$, and $\zeta\in (-\zeta_{1},\zeta_{1})$. The
function $\partial_{x_0}x(t;\hat{x},0)$ satisfies the linear
equation
\begin{equation*}
\text{d}_t\partial_{x_0}x(t;\hat{x},0)=\left.\partial_{x_0}
\f(x;\zeta)\right|_{(\hat{x},0)}=\partial_{x}\f(\hat{x};0)^{\top}\partial_{x_0}
x(t;\hat{x},0), \quad \partial_{x_0} x(0;\hat{x},0)=\mathds{1}.
\end{equation*}

Further linear equation for $\partial_{\zeta}x$ can also be derived
to show that continuity of $\f$ gives continuous differentiability
of $S$ on $\R^2\times (-\zeta_1,\zeta_1)$. It then holds
$\partial_{x_0}S(\hat{x},0)=\partial_{x_0}x(\omega^{-1};\hat{x},0)-\mathds{1}=\exp(\partial_{x}\f(\hat{x};0)\omega^{-1})-\mathds{1}$.
We want to know under which condition this matrix has a bounded
inverse. Let $v$ be an eigenvector of $\partial_{x}\f(\hat{x};0)$
that associates with an eigenvalue $\lambda$. The Taylor expansion
for matrix exponential gives us
$(\exp(\partial_{x}\f(\hat{x};0)\omega^{-1})-\mathds{1})
v=(\exp(\omega^{-1}\lambda)-1) v$, making
$\exp(\omega^{-1}\lambda)-1$ the associated eigenvalue of
$\partial_{x_0}S(\hat{x},0)$. It remains to show that for any
eigenvalue $\lambda$ of the Jacobian $\partial_{x}\f(\hat{x};0)$,
$\exp(\omega^{-1}\lambda)-1$ can never be zero or
\begin{equation}\label{eq:check}
\lambda\neq 2\pi\omega i\mathbb{Z}
\end{equation}
where $i,\mathbb{Z}$ denote the imaginary number and the set of
integers, respectively.

One sufficient condition for the invertibility of
$\partial_{x_0}S(\hat{x},0)$ is to have a negative trace of the
Jacobian. By the Implicit Function Theorem, there exist a domain
$U_2(\hat{x})\times(-\zeta_{2},\zeta_{2})$ and a continuously
differentiable function $x_0(\zeta)$ for which $(\zeta, x_0(\zeta))$
is defined on this domain such that $S(x_0(\zeta),\zeta)=0$ or
eventually $ x(\omega^{-1}, x_0(\zeta),\zeta)= x_0(\zeta)$. Since
$\f$ is $\omega^{-1}$--periodic over time, then $x(t+\omega^{-1},
x_0(\zeta),\zeta)= x(t, x_0(\zeta),\zeta)$ if and only if
$x(\omega^{-1}, x_0(\zeta),\zeta)= x_0(\zeta)$. The desired domain
for $(x_0,\zeta)$ for the existence of the $\omega^{-1}$--periodic
function can then be restricted to $\left\{U_1(\hat{x})\cap
U_2(\hat{x})\right\}\times
\left\{(-\zeta_{1},\zeta_{1})\cap(-\zeta_{2},\zeta_{2})\right\}$.

Note that such a negative trace of the Jacobian would give a
dissipative autonomous system with the exponential dissipative rate
given by the trace, which is another way to see that the autonomous
system contracts to a set of measure zero. Added by
Dulac--Bendixson's criterion, the negative trace naturally
guarantees non-existence of a limit cycle in the nonnegative
quadrant for the autonomous system. When seasonal forcing is
activated, this condition prevents the birth of a trajectory where a
limit cycle is interfered by such seasonal forcing, i.e.\ a torus.

\subsection{Approximation of periodic solutions}\label{sec:approx}
Here we center the investigation on what would be the behaviour of
the periodic solutions under variation of the amplitude $\zeta$ and
adiabatic parameter $\epsilon$. The sinusoidal function $\phi=\zeta
h$ where $h$ is as in \eqref{eq:h} can be seen as the solution of
the differential equation $(\phi',\psi')=(\psi,-4\pi^2\omega^2\phi)$
where $(\phi,\psi)(0)=(\zeta A_c\slash \bar{M},2\zeta
A_s\pi\omega\slash \bar{M})$. Under the autonomous setting, the
entire system decouples into
\begin{equation}\label{eq:full}
\begin{split}
j'&=\beta(j)\cdot(1-j)v-\kappa
j,\\
v'&=\rho(1+\phi-v)j-\theta v,\\
\phi'&=\psi,\\
\psi'&=-4\pi^2\omega^2\phi.
\end{split}
\end{equation}
Let $(\hat{j},\hat{v})$ denote an equilibrium state of
$(j,v)$--dynamics. The only equilibrium state of
$(\phi,\psi)$--dynamics is $(\hat{\phi},\hat{\psi})=(0,0)$ with
purely imaginary eigenvalues, leading to the fact that any
equilibrium of \eqref{eq:full} is non-hyperbolic. As an intermediate, let us recall the numerical
estimates for observable parameters from Tab.~\ref{tab:par} to get
an idea of how large $\phi,\psi$ would be. We apparently obtain
$$
\text{Ampl}[\phi]=\zeta\cdot\text{Ampl}[h]=\left(\frac{\mu}{\xi\sqrt{\theta^2+4\pi^2\omega^2}}\right)\left(\frac{\zeta}{\epsilon}\right)\quad\text{and}\quad\text{Ampl}[\psi]=2\pi\omega\cdot
\text{Ampl}[\phi].
$$
We also observe from the estimates that
$\text{Ampl}[\phi],\text{Ampl}[\psi]\sim \zeta\slash\epsilon$.
Providing that $\epsilon>\mu\zeta\slash\xi\sqrt{\theta^2+4\pi^2\omega^2}$, we can employ Center
Manifold Theorem to have a realization of $(j,v)$--dynamics in the
center manifold given by
\begin{align*}
j(\phi,\psi)&=\hat{j}+a_1\phi^2+a_2\phi\psi+a_3\psi^2+\OM(\lVert (\phi,\psi)\rVert^3),\\
v(\phi,\psi)&=\hat{v}+b_1\phi^2+b_2\phi\psi+b_3\psi^2+\OM(\lVert
(\phi,\psi)\rVert^3).
\end{align*}
One can naturally neglect the third-order terms by the preceding
estimates of $\phi,\psi$. It remains to determine the constants
appearing in the ansatzs by, instead of proximity optimization,
equating $j',v'$ from the actual dynamics and the ansatzs as well as
focusing solely on small-order terms:
\begin{align*}
j'&=(2a_1\phi+a_2\psi)\psi+(2a_3\psi+a_2\phi)(-4\pi^2\omega^2\phi)=\left.\beta(j)\cdot(1-j)v-\kappa j\right|_{j(\phi,\psi),v(\phi,\psi)},\\
v'&=(2b_1\phi+b_2\psi)\psi+(2b_3\psi+b_2\phi)(-4\pi^2\omega^2\phi)=\left.\rho(1+\phi-v)j-\theta
v\right|_{j(\phi,\psi),v(\phi,\psi)}.
\end{align*}
Depending on how complicated $\beta$ is, the computations of the
constants can get laborious. Prominent from this investigation is
that the periodic ansatzs of $j,v$ in the center manifold perturb
from the equilibrium states $\hat{j},\hat{v}$ with sinusoidal
functions of amplitudes equivalent to $(\zeta\slash\epsilon)^2$ in
case $\epsilon>\mu\zeta\slash\xi\sqrt{\theta^2+4\pi^2\omega^2}$. In case $\epsilon\leq
\mu\zeta\slash\xi\sqrt{\theta^2+4\pi^2\omega^2}$, the higher orders in the center manifold
approximation matter and $\epsilon$ takes the lead in extending the
amplitude. Therefore, no small-order polynomial approximation of the periodic
solutions can be envisaged. A crude estimate on the vector lifespan
$\theta^{-1}\approx 4\text{w}$ gives $\epsilon=\mu\slash\theta
=4\mu\geq 4\mu\zeta\slash\xi> \mu\zeta\slash\xi\sqrt{\theta^2+4\pi^2\omega^2}\approx 3.6\mu\zeta\slash\xi$, suggesting that the second-order
approximation is already quite reasonable. The preceding exposition gives us a new
tool in designing the model to have periodic solutions that will
correct the assimilation to given data. In Section~\ref{sec:asump},
the solution of the autonomous model was deduced to portray lightly fluctiative data
irrespective of $\epsilon$. In case the seasonal forcing is
activated, we can make use of both $\zeta$ and $\epsilon$ to create
an amplitude that delineates that of the data. Questionable is thus
how to correctly specify the numerical value of $\rho_{\epsilon}$.
We have just foreseen the use of the model for data assimilation
under specification of three unobservable parameters.

\section{When hospitals can accommodate an unlimited number of patients}\label{sec:report}
Note, first of all, that both $j$ and $I$ can be treated equally in
the modeling since $j$ is linearly proportional to $I$. In such
case, we can directly initiate models of the infection rate $\beta$ as a function of
$j$. Suppose that in the absence of disease, $j=0$, the introduction
of infective vector population gives rise to a certain driving
force, represented by the initial infection rate $\beta_0$. We suppose that $\beta$ stays
more-or-less constant while the hosts are still unaware of the
ongoing infections, then initiates decrement around a certain
reference point $j^{\ast}\in [0,1)$. This threshold $j^{\ast}$
indicates the alarming incidence level under which medical
departments can never transfer information to media holders for
either time, interest or financial restrictions. The interval on
which such driving force decreases, i.e.\ $j^{\ast}\leq j$, is that
when the humans stay alarmed of the ever increasing infections,
i.e.\ where self-precautions and hospitalizations are urgently
undertaken. Of course, such awareness can never be achieved unless
the media keep reporting on current infection cases. When $j$
increases further, a slow downturn in the infection rate is
assigned due to fewer contacts between hosts and vectors. When $j$
is close to the maximum ($j=1$), $\beta$ saturates to a certain
level $\beta_{1}<\beta_0$, since most of the hosts are aware of the
danger and to be equipped with uniform self-defence system, also
many of them are hospitalized. The preceding description leads us to
the following summary on $\beta$:
\begin{enumerate}
\item[(A1)] $\beta(j)>0$ for all $j\in [0,1]$,\label{item:A1}
\item[(A2)] $\beta_0=\beta(0)>\beta(1)=\beta_1$,
\item[(A3)] $\beta'(j)=0$ for $j\in [0,j^{\ast}]$ and $\beta'(j)<0$ for all $j\in (j^{\ast},1]$.\label{item:A3}
\end{enumerate}

The slow dynamics $j$ in the critical manifold are governed by
\begin{equation}\label{eq:slow}
j'=J_0(j)+J_1(j)v^{\ast}(j)=\beta(j)\cdot(1-j)\frac{\rho_{\epsilon}j}{\rho_{\epsilon}j+\mu}-\kappa
j.
\end{equation}
We know that $[0,1]$ is invariant due to the boundary conditions
$j'(0)=0$ and $j'(1)<0$. At the end of Section~\ref{sec:slowfast},
we came into understanding that whichever equilibrium of
\eqref{eq:slow} should be where the critical and slow
manifold intersect. The equation \eqref{eq:slow} has the trivial equilibrium,
known as \emph{disease-free equilibrium}, $j=0$. It is instantly
verifiable that the Jacobian
$\left.\text{d}_{j}j'\right|_{j=0}=\beta_0\rho_{\epsilon}\slash
\mu-\kappa=\kappa(\beta_0\rho_{\epsilon}\slash \mu \kappa - 1)<0$,
providing that the \emph{basic reproductive number}
\begin{equation}\label{eq:R0}
\RM_0:=\sqrt{\beta_0\rho_{\epsilon}\slash \mu \kappa}<1.
\end{equation}
At this point, we acquire local asymptotic stability of $j=0$. If
$\RM_0>1$, then $j=0$ is unstable. If $\RM_0=1$, then one can take a
small $\varepsilon>0$ and easily verify that
$\left.j'j\right|_{j=\varepsilon}=-(\kappa+\beta(\varepsilon))\varepsilon^2+\beta(\varepsilon)\varepsilon>0$
due to $\lim_{\varepsilon\rightarrow
0^+}\beta(\varepsilon)\varepsilon\slash
(\kappa+\beta(\varepsilon))\varepsilon^2=\beta_0\slash
(\kappa+\beta_0)\lim_{\varepsilon\rightarrow 0^+}1\slash
\varepsilon=\infty$. This indicates that $j=0$ is repelling in the
positive real.

Calculating a nontrivial solution, i.e.\ \emph{endemic equilibrium},
is quite straightforward. Factoring out $j$ from \eqref{eq:slow}, we
are in the position to solve
\begin{equation}\label{eq:nontrivial}
E(j):=\beta(j)\cdot (1-j)\rho_{\epsilon}-\kappa
(\rho_{\epsilon}j+\mu)=0.
\end{equation}
Observe that
$E(0)=\beta_0\rho_{\epsilon}-\mu\kappa=\mu\kappa(\mathcal{R}_0^2-1)$,
$E(1)=-\kappa (\rho_{\epsilon}+\mu)<0$ and $E'(j)=\beta'(j)\cdot
(1-j)\rho_{\epsilon}-\beta(j)\cdot\rho_{\epsilon}-\kappa\rho_{\epsilon}<0$
for all $j\in(0,1]$ due to (A1)--(A3). We acquire a dichotomy. If
$\mathcal{R}_0\leq 1$, then there exists no endemic equilibrium. If
$\mathcal{R}_0>1$, then there exists a unique endemic equilibrium
$j=j_e\in (0,1)$ approaching the disease-free equilibrium as
$\mathcal{R}_0\rightarrow 1^{+}$, i.e.\ due to fixed $E(1)$.
Furthermore, as we calculate the Jacobian, it turns out to be
\begin{equation}\label{eq:Ep}
\left.\text{d}_jj'\right|_{j=j_e}=\left.\text{d}_j\frac{E(j)\cdot
j}{(\rho_{\epsilon}j+\mu)}\right|_{j=j_e}=\left.\frac{E'(j)\cdot
j}{(\rho_{\epsilon}j+\mu)} + E(j)\cdot
\frac{\mu}{(\rho_{\epsilon}j+\mu)^2}
\right|_{j=j_e}=\frac{E'(j_e)\cdot j_e}{(\rho_{\epsilon}j_e+\mu)}
\end{equation}
such that
\begin{equation}\label{eq:sgn}
\text{sign}\left(\left.\text{d}_jj'\right|_{j=j_e}\right)=\text{sign}(E'(j_e)).
\end{equation}
Additionally attributed to $\mathcal{R}_0>1$ is thus the local
asymptotic stability of the endemic equilibrium. After all, the preceding
exposition shows that the slow dynamics using $\beta$ fulfilling
(A1)--(A3) exhibits a supercritical bifurcation at
$\mathcal{R}_0=1$. Note that the bifurcation profile cannot change
with the adiabatic parameter $\epsilon$ due to
$\mathcal{R}_0$'s independency on it. Moreover, the similar
bifurcation would appear in case $\beta$ is strictly monotonically
decreasing. This stems from the simple fact that the analysis is
independent on $j^{\ast}$, i.e.\ that one can thus set $j^{\ast}=0$
in (A3). Similar results of supercritical bifurcation with respect
to $\mathcal{R}_0$ using non-increasing $\beta$ can also be seen in
\cite{LC2008}.

As $\epsilon>0$, we still keep the full system \eqref{eq:jv} on our
hand. Apparently, the Jacobians evaluated at the disease-free
equilibrium $(0,0)$ and endemic equilibrium $(j_e,v_e)$ with
$v_e=v^{\ast}(j_e)$ have the traces
$$
-\beta_0v_e-\kappa-\theta\quad\text{and}\quad
\beta'(j_e)\cdot(1-j_e)v_e-\beta(j_e)\cdot v_e- \kappa-\rho
j_e-\theta
$$
respectively, which are clearly negative. One can thus not expect to
have purely imaginary eigenvalues for each Jacobian, fulfilling the
existence check \eqref{eq:check}. A closing statement in this
section is that there always exist $\omega^{-1}$--periodic solutions
surrounding the equilibria.

\section{Limited medical access and the increase of infection rate}\label{sec:nonreport}
Here we consider a scenario where the available hospitals in the
observed region can only accommodate a certain population share
$j^{\ast}\in [0,1)$. As such, the alarming incidence level
$j^{\ast}$ gains a new definition. We consider a setting where no
reports to the media are initiated in case hospitals still manage to
shelter patients. Unawareness of the susceptible hosts for the case
$j\leq j^{\ast}$ thus leads to a constant $\beta=\beta_0$. A
decrement in $\beta$ happens when the media keep reporting not only the
current endemicity, but also the absence of medical access. The phase, in which case a sudden
decrement in the infection rate happens, is what we consider as that when susceptible hosts
\emph{overreact} to the advancing disease propagation and that there
is nowhere to go for cure.

\begin{figure}[htbp!]
\centering \psfrag{j}{\large$j$}\psfrag{beta}{\large$\beta(j)$}
\psfrag{jst}{$j^{*}$}\psfrag{c}{$c$}\psfrag{B0}{$\beta_{0}$}
\includegraphics[width=0.55\textwidth]{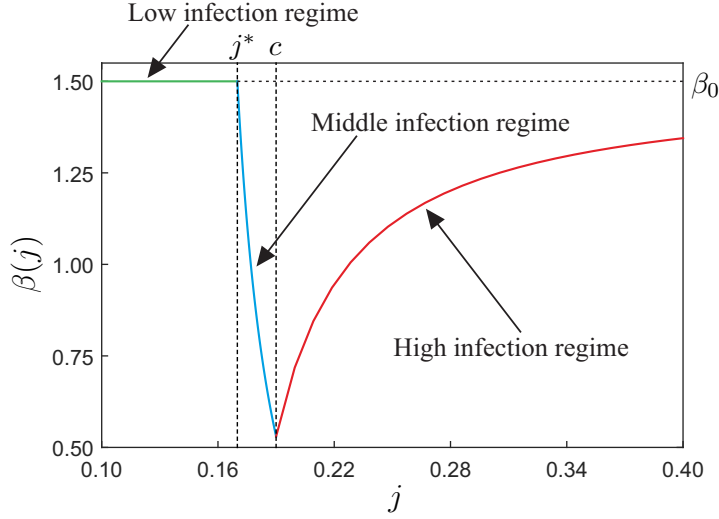}
\caption{Behavior of the infection rate considering media reporting
and several factors persuading increment for larger $j$. The curve
is computed for $a=0.53$, $\beta_{0}=1.5$, $j^{*}=0.17$ and
$c=0.19$, see \eqref{eq:newbeta}. The infection rate defines three
operation modes: low infection regime ($j\leq j^{*}$), middle
infection regime ($j^{*}< j\leq c$) and high infection regime
($j>c$).}\label{fig-beta-curve}
\end{figure}

Let us now consider a situation where the investigated vector-borne
disease is endemic in a developing region. By developing we point at
the human tendency of disobeying regulations towards healthy life
and inability as well as unwillingness to regularly afford
household's preventive medicines. It is also ubiquitous that developing regions be identical with ``population explosion'' and limited medical access. The fact that vector-borne
diseases, such as dengue, are endemic in areas of relatively hot
temperature and high air humidity that befit well with vector
breeding, the concepts of putting on full-cover clothing and
smearing repellent fluids sound counter-intuitive. Other measures
such as disseminating \emph{temephos}, cleaning household vessels,
and fumigation usually go on individual basis and require a big
campaign for the effectivity in a population scale \cite{SNP2018}. The fact
that disease vectors, such as \emph{Aedes aegypti} mosquitoes, only
bite during daylight also makes such small self-defence measures
prone to discontinuation. Despite high awareness about the
disease and preventive measures,
a study has shown that typical vector-borne disease with a low case-fatality
rate is perceived deadly by only a portion of susceptible humans,
while others perceive it easily curable \cite{WB2013}. Here we consider
the situation where $\beta$ revolves in the likelihood of disowning
preventive measures, absence of hospitals, poverty, and perceiving
the disease of being easily curable. A prominent psychological
trigger can be seeing through increasing incidences in the
neighbourhood with despair, as continuous usage of preventive
measures can seemingly not suppress the incidence level. Another
unforeseeable phenomenon when $j$ increases further, such that the
estimate of mosquito population $v^{\ast}(j)$ also increases, is
that the susceptible humans are evermore surrounded by hungrier
vectors \cite{SRD2012}. This surely gives additional correction to $\beta$
for sufficiently large $j$. In summary, we consider $\beta$ that
satisfies:
\begin{enumerate}
\item[(B1)] $\beta(j)>0$ for all $j\in [0,1]$,\label{item:B1}
\item[(B2)] $\beta'(j)=0$ for $j\in [0,j^{\ast}]$ with $\beta(0)=\beta_0$ and $\beta'(j)<0$ for all $j\in (j^{\ast},c]$ for a certain positive $c<1$,\label{item:B2}
\item[(B3)] $\beta'(j)>0$ for all $j\in (c,1]$ and slowly saturates to $\beta(1)=\beta_1$.\label{item:B3}
\end{enumerate}%
Based on the preceding specifications, we propose the following
ansatz
\begin{equation*}
\beta(j):=a+\frac{b|j-c|}{d+|j-j^{\ast}|}
\end{equation*}
for some constants $a,b,c,d$. Relying on (B2), we can compute $b,d$
to have
\begin{equation}\label{eq:newbeta}
\beta(j):=a+\frac{(\beta_0-a)|j-c|}{c-j^{\ast}+|j-j^{\ast}|}.
\end{equation}
This formula clearly produces spikes at $j^{\ast}$ and $c$. The
parameter $a$ represents the minimum value of $\beta$, which is
attained at $c$. This definition also means $\beta_0\geq a$.
Moreover, there are several ways to treat $\beta$ as to study
bifurcation. The basic idea is to keep some parameters fixed, while
others vary. Since we look further upon the variation of equilibria
with respect to that of the basic reproductive number, varying
$\beta_0$ shall do the job. We additionally assume that $a$,
$j^{\ast}$ and $c$ are observable, leading to the variation of
$\beta_1$ as $\beta_0$ varies. This way also provides flexibility in
to which level $\beta$ ultimately increases when $j$ gets larger.
Further analysis also shows that $1-c<1-j^{\ast}$, leading to
$\beta_1<\beta_0$ for all possible choices of $\beta_0$. Finally,
the realization of $\beta$ for a certain set of parameters can be
seen in Fig.~\ref{fig-beta-curve}. There we have classified the
domain $[0,1]$ into three regimes based on the value of the
incidence level $j$.

As far as the slow dynamics $j$ in the critical manifold is
concerned, we still get from \eqref{eq:slow} using $\beta$ in
\eqref{eq:newbeta} the disease-free equilibrium $j=0$. The same
investigation over the Jacobian $\text{d}_jj'$ shows local
asymptotic stability of the equilibrium in case $\mathcal{R}_0<1$
and instability in case $\mathcal{R}_0\geq 1$.

Now let us analyze the existence of endemic equilibria. The most
important insights from the formulation of $\beta$ in
\eqref{eq:newbeta} are that $\beta\rightarrow \beta_0$ as
$j^{\ast}\rightarrow c^-$ and that $\beta<\beta_0$ as $c>j^{\ast}$.
Let us recall the function $E$ in \eqref{eq:nontrivial} for the
purpose of finding endemic equilibria. The entities
$E(0)=\mu\kappa(\mathcal{R}^2_0-1)$ and
$E(1)=-\kappa(\rho_{\epsilon}+\mu)<0$ remain unchanged. The
aforementioned insights further reveal that $E$ is bounded above by
a linear function in $j$, i.e.\
$$
E(j)\leq
\beta_0(1-j)\rho_{\epsilon}-\kappa(\rho_{\epsilon}j+\mu)=-(\beta_0\rho_{\epsilon}+\kappa\rho_{\epsilon})j+E(0)\quad\text{for
all }j\in [0,1],
$$
where the equality holds for $c=j^{\ast}$. With this finding, the
existence analysis becomes much simple. If $\mathcal{R}_0\leq 1$,
then $E(0)\leq 0$ and $E$ can never have roots in $(0,1]$, therefore
no endemic equilibrium is prevailing. For the sake of further
analysis, let us reveal
\begin{equation}\label{eq:Ej}
E(j^{\ast})=\frac{\rho_{\epsilon}(2+c-3j^{\ast})}{c-j^{\ast}}\beta_0-\rho_{\epsilon}\frac{2a(1-j^{\ast})}{c-j^{\ast}}-\rho_{\epsilon}(\kappa+2a).
\end{equation}
If $\mathcal{R}_0>1$, we can consider two cases: $E(j^{\ast})\leq 0$
and $E(j^{\ast})>0$, each of which is dependent on $\beta_0$. The
former gives the existence of a unique endemic equilibrium in
$(0,j^{\ast}]$, which is locally asymptotically stable due to
$E'<0,E=0$ at that point, see \eqref{eq:sgn}. The latter case
$E(j^{\ast})>0$ requires information on
$E(c)=a(1-c)\rho_{\epsilon}-\kappa(c\rho_{\epsilon}+\mu)$, which is
surely greater than $E(1)$. When $E(c)>0$, we have the existence of
a unique endemic equilibrium in $(c,1)$, which is also locally
asymptotically stable. When $E(c)=0$, we need to check the shape of
$E$ on $(c,1]$ in order to estimate the other endemic equilibria
therein. We found that
\begin{align}
E'(c^+)&=\frac{\rho_{\epsilon}(1-c)}{2(c-j^{\ast})}\beta_0-a\rho_{\epsilon}\frac{1-j^{\ast}}{2(c-j^{\ast})}-\rho_{\epsilon}(a+2\kappa),\label{eq:Epc}\\
E''(j)&=\frac{-4\rho(\beta_0-a)(c-j^{\ast})(1+c-2j^{\ast})}{(j+c-2j^{\ast})^3}<0\quad\text{for
all }j\in (c,1]\label{eq:Epp}
\end{align}
in case $c>j^{\ast}$. The second result indicates that $E$ is concave
on $(c,1]$. We see that two endemic equilibria can exist in case
$\beta_0$ makes $E'(c^+)>0$. Suppose that this is indeed the case.
We naturally lose the information regarding the stability of $j=c$
due to non-uniqueness of $E'(c)$ due to the non-unique subgradients. The
other equilibrium in $(c,1)$ surely is locally asymptotically stable
because of $E'<0$.

The last case $E(c)<0$ gives a unique endemic equilibrium in
$(j^{\ast},c)$, which is, again, locally asymptotically stable due
to \eqref{eq:sgn}. From \eqref{eq:Ej} and \eqref{eq:Epc}, we come to
understand that $\beta_0$ acts to ``pull'' the curve of $E$ towards
positivity on $(c,1]$, until then the curve delineates the
decreasing straight line as $E$ devolves. Therefore, depending on
how close $c$ and $j^{\ast}$ are and how large $\beta_0$ is, we can
have either none, one, or two more endemic equilibria in $(c,1)$. In
case one endemic equilibrium is found in $(c,1)$, it should be
equipped with $E=E'=0$, which leads us to unknown status regarding
its stability. In case two endemic equilibria are found, it thus is
obvious that the smaller one is unstable, while the larger one is
locally asymptotically stable due to \eqref{eq:sgn}. Another
obviousness is that the equilibrium in $(j^{\ast},c)$ and the
smaller equilibrium in $(c,1)$ become closer to each other as
$j^{\ast}$ and $c$ get closer. As far as $E(c)<0$ is concerned, we
can derive a sufficient condition such as $a\leq\kappa c\slash
(1-c)$ that will lead to it, where more certainty can be gained as
$c$ walks towards $1$.

We are now analyzing the existence of periodic solutions surrounding
the existing equilibria. Suppose that $\zeta>0$ and $\epsilon>0$.
Let $(\hat{j},\hat{v})$ be an equilibrium of the full system where
$\hat{v}=v^{\ast}(\hat{j})$. The Jacobian of the system evaluated at
the equilibrium in terms of the function $E$ takes the form
\begin{equation}\label{eq:Jac}
\partial_{x}\f(\hat{x};0)=\left(\begin{array}{cc}
\frac{1}{\rho_{\epsilon}\hat{j}+\mu}\left(E'(\hat{j})\hat{j}-\kappa\mu\right)& \beta(\hat{j})\cdot(1-\hat{j})\\
\rho(1-\hat{v})&-\rho\hat{j}-\theta
\end{array}
\right).
\end{equation}
Suppose that $\beta_0$ is large enough such that four equilibria are
present. When $\hat{j}=0$, one can easily verify that the trace of
the Jacobian is negative, avoiding having purely imaginary
eigenvalues. Therefore, a $\omega^{-1}$--periodic solution
surrounding the disease-free equilibrium $(0,0)$ exists, which turns
to be the equilibrium itself. One can easily verify this by the fact that the
disease-free equilibrium solves the non-autonomous model
irrespective to the values of $\zeta$. The endemic equilibria
corresponding to the smallest and largest $\hat{j}$ were shown to
fulfill $E'(\hat{j})<0$, making the trace of the Jacobian negative.
This evidence, once again, shows that $\omega^{-1}$--periodic
solutions surrounding the two endemic equilibria exist. The
equilibrium corresponding to $\hat{j}$ in between the other two
endemic equilibria discovers $E'(\hat{j})>0$. Variation in
$E'(\hat{j})\hat{j}-\kappa\mu$ can thus make the trace obtain either
a negative, zero, or a positive value. In case the trace is zero,
then $E'(\hat{j})\hat{j}-\kappa\mu$ must have been positively large
enough, but then the determinant of the Jacobian becomes negative.
We acquire two real roots of the same absolute value that solely
oppose in sign. Therefore, a $\omega^{-1}$--periodic solution
surrounding this middle endemic equilibrium also exists.

\section{Numerical analysis of the model via path-following methods}

The epidemic model considering the infection rate introduced in the
previous section can be studied in the framework of
\emph{piecewise-smooth dynamical systems} \cite{BBC2004}. This
type of systems arises typically when some kind of nonsmooth
phenomenon is considered, such as (soft) impacts, switches,
friction, etc. The system response in this case is determined by a
piecewise-smooth vector field due to the presence of discrete events
producing discontinuities in first or higher-order derivatives of
the solution. In our epidemic model, the nonsmoothness is produced
by sharp transitions in the force of infection due to the social
behavior towards media reports as well as poverty and reluctance in
applying measures against the spread of the disease.

In general, a piecewise-smooth dynamical system can be defined in
terms of two main components: a collection of (smooth) vector fields
and \emph{event functions}. In this way, the system is characterized
by a number of operation modes, each of which is associated with a
specific smooth vector field. On the other hand, the event functions
define the boundary for the operation modes, in such a way that
whenever the solution crosses transversally a certain boundary
defined by (usually the zero-set of) an event function, the system
changes to a different operation mode governed by a (possibly)
different vector field. In this way, any solution of the
piecewise-smooth dynamical system can be represented by a sequence
of \emph{segments}, which consists of a pair given by a smooth
vector field describing the model behavior and an event function
that defines the terminal condition for the operation mode. More
details about this formulation can be found in
\cite{TD2008,DS2013}.

\subsection{Mathematical setup}\label{sec:pws}

For the numerical study of the underlying epidemic model via
continuation methods, it is convenient to write the governing
equations in autonomous form. To do so, we will consider the
following nonlinear oscillator that will be appended to the system
\cite{KOG2007}:
\begin{equation}\label{eq-nonosc}
\begin{cases}
p'=p+2\pi\omega q-p\left(p^2+q^2\right),\\
q'=q-2\pi\omega p-q\left(p^2+q^2\right),
\end{cases}
\end{equation}
with the asymptotically stable solution $p(t)=\sin(2\pi\omega t)$
and $q(t)=\cos(2\pi\omega t)$. In this way, we can write the
periodically forced system \eqref{eq:jv} in the autonomous setting,
which then allows us to study the model via numerical continuation
methods. Let us define
$\alpha:=(\beta_{0},a,j^{*},c,\kappa,\mu,\epsilon,\rho_{\epsilon},\xi,\omega,\zeta)\in\left(\R^+\right)^{10}\times\R^+_{0}$
and $z(t):=(j(t),v(t),p(t),q(t))\in\left(\R^+_{0}\right)^{2}\times\R^2$ as the
parameters and state variables of the system, respectively, where
$\R^+_{0}$ stands for the set of nonnegative real numbers. As
explained above, any solution of the considered Dengue epidemic
model can be divided into the following segments (see also Fig.\
\ref{fig-beta-curve}).

\textbf{Low infection regime}. This segment occurs when the
incidence level is low (i.e.\ $j\leq j^{*}$). Therefore, the media
do not apprise susceptible hosts of such a state. Emanating from
\eqref{eq:jv} coupled with the seasonal forcing, the model behavior
during this regime is governed by the (smooth) ordinary differential
equation
\begin{equation}\label{eq-LI-ode}
z'=\f_{\text{\tiny
LI}}(z,\alpha):=\left(\setstretch{1.5}\begin{array}{c}
\beta_{0}(1-j)v-\kappa j\\
\dfrac{\rho_{\epsilon}}{\epsilon}\left(1+\frac{\zeta A_c}{\bar{M}}q+\frac{\zeta A_s}{\bar{M}}p-v\right)j-\dfrac{\mu}{\epsilon}v\\
p+2\pi\omega q-p\left(p^2+q^2\right)\\
q-2\pi\omega p-q\left(p^2+q^2\right)
\end{array}\right).
\end{equation}
This segment terminates when the incidence level 
increases beyond $j^{*}$, which can be detected via the \emph{event
function} $h_{1}(z(t),\alpha):=j(t)-j^{*}=0$. In this case, the
system switches to the operation regime given below.

\textbf{Middle infection regime}. During this operation mode, the
incidence level lies in the window $j^{*}< j\leq c$, where the
hospitals cannot accept more patients and appeal for regular reports
eventually made by the media. As a result of \emph{overreaction},
the rest of susceptible hosts forcefully mobilize everything to keep
them on the safe zone. This has then the effect of decreasing the
force of infection, thus eventually decreasing the infection rate
(see Fig.\ \ref{fig-beta-curve}). During this regime, the model
obeys the following equation
\begin{equation}\label{eq-MI-ode}
z'=\f_{\text{\tiny
MI}}(z,\alpha):=\left(\setstretch{1.5}\begin{array}{c}
\left(a-\dfrac{(\beta_{0}-a)(j-c)}{j+c-2j^{*}}\right)(1-j)v-\kappa j\\
\dfrac{\rho_{\epsilon}}{\epsilon}\left(1+\frac{\zeta A_c}{\bar{M}}q+\frac{\zeta A_s}{\bar{M}}p-v\right)j-\dfrac{\mu}{\epsilon}v\\
p+2\pi\omega q-p\left(p^2+q^2\right)\\
q-2\pi\omega p-q\left(p^2+q^2\right)
\end{array}\right).
\end{equation}
This regime terminates in two ways. First, the incidence level
decreases below $j^{*}$, in which case $h_{1}(z(t),\alpha)=0$.
Consequently, the system switches back to the low infection regime
defined above. Second, the incidence level increases beyond $c$,
which can be detected via the vanishing \emph{event function}
$h_{2}(z(t),\alpha):=j(t)-c=0$. If this occurs, then the system
operates under the regime defined next.

\textbf{High infection regime}. This regime certifies the increasing
infection rate with the incidence level $j>c$ (see Fig.\
\ref{fig-beta-curve}). The hypothetical causes were due to poverty,
hungrier vectors, absence of medical access, reluctancy in regularly
taking up preventive measures, and despair caused by long lasting
high levels of incidences despite keeping up caution. The model
behavior during this regime then obeys the equation
\begin{equation}\label{eq-HI-ode}
z'=\f_{\text{\tiny
HI}}(z,\alpha):=\left(\setstretch{1.5}\begin{array}{c}
\left(a+\dfrac{(\beta_{0}-a)(j-c)}{j+c-2j^{*}}\right)(1-j)v-\kappa j\\
\dfrac{\rho_{\epsilon}}{\epsilon}\left(1+\frac{\zeta A_c}{\bar{M}}q+\frac{\zeta A_s}{\bar{M}}p-v\right)j-\dfrac{\mu}{\epsilon}v\\
p+2\pi\omega q-p\left(p^2+q^2\right)\\
q-2\pi\omega p-q\left(p^2+q^2\right)
\end{array}\right).
\end{equation}
This segment terminates when the level of incidences decreases below
$c$ (i.e.\ $h_{2}(z(t),\alpha)=0$), in such a way that the system
operates then under the middle infection regime defined above.

Under this setting, the epidemic model introduced in the previous
section can be written as a piecewise-smooth dynamical system as
follows
\begin{equation}\label{eq-pws}
\begin{cases}
z'=\f_{\text{\tiny LI}}(z,\alpha), & j\leq j^{*}\mbox{ }\mbox{ }\mbox{ }(\mbox{low infection regime}),\\
z'=\f_{\text{\tiny MI}}(z,\alpha), & j^{*}< j\leq c\mbox{
}\mbox{ }\mbox{ }(\mbox{middle infection regime}),\\
z'=\f_{\text{\tiny HI}}(z,\alpha), & j>c\mbox{ }\mbox{ }\mbox{
}(\mbox{high infection regime}).
\end{cases}
\end{equation}

\subsection{Numerical investigation of the epidemic model
subject to one-parameter variations}

In this section, our main goal is to study the behavior of the model
\eqref{eq-pws} when selected parameters are varied. For this
purpose, we introduce a solution measure in order to interpret the
numerical results in the context of the considered epidemiological
scenario. Suppose that $(j,v,p,q)$ is a bounded periodic solution of
system \eqref{eq-pws} with the fundamental period $\omega^{-1}$.
Under this assumption, we define the following solution measure
\begin{equation}\label{eq-jmax}
j_{\mbox{\scriptsize max}}:=\max_{0\leq t\leq \omega^{-1}}j(t),
\end{equation}
which gives the peak of incidence level within the time window
$[0,\omega^{-1}]$. Therefore, one of the main concerns is to
investigate under what conditions $j_{\mbox{\scriptsize max}}$ can
be kept as low as possible. In addition, we consider the solution
measure
\begin{equation}\label{eq-jP}
j_{\mbox{\scriptsize p}}:=\max_{0\leq t\leq
\omega^{-1}}j(t)-\min_{0\leq t\leq \omega^{-1}}j(t),
\end{equation}
which gives the peak-to-peak amplitude of the $j$-component of the
periodic solution.

\begin{figure}[htbp!]
\centering
\psfrag{a}{\large(a)}\psfrag{b}{\large(b)}\psfrag{J1}{\large$j$}
\psfrag{V1}{\large$v$}\psfrag{D}{\Large$j$}\psfrag{E}{\Large$v$}
\psfrag{MJ}{\Large$j_{\mbox{\scriptsize
max}}$}\psfrag{zeta}{\Large$\zeta$}\psfrag{jst}{\large$j^{*}$}\psfrag{c}{\large$c$}
\includegraphics[width=\textwidth]{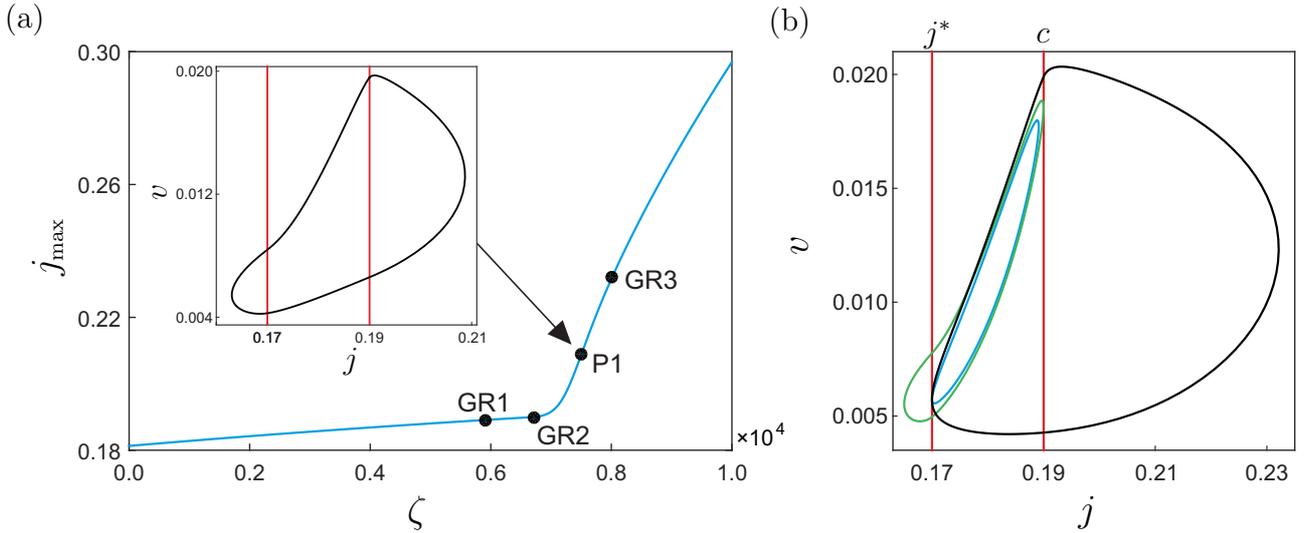}
\caption{One-parameter continuation of the periodic response shown
in panel (a) (inner diagram) with respect to the amplitude of
seasonality effects $\zeta$, for the parameter values given in
Table~\ref{tab:par}, with $a=0.53$, $\beta_{0}=1.5$, $j^{*}=0.17$,
$c=0.19$ and $\epsilon=9.8\times10^{-4}$. The left panel depicts the
behavior of the peak of incidence level $j_{\mbox{\scriptsize max}}$
(see \eqref{eq-jmax}) as the parameter $\zeta$ varies. The labels
GR$i$ stand for grazing solutions shown in panel (b) (GR1 in blue,
GR2 in green, GR3 in black), while the point P1 marks the initial
periodic solution taken at $\zeta=7500$. Vertical red lines in the
phase plots stand for the discontinuity boundaries defined at
$j=j^{*}$ and $j=c$.}\label{fig-bif-diag-zeta}
\end{figure}

To analyze the behavior of the model \eqref{eq-pws}, we employ
path-following (numerical continuation) methods for piecewise-smooth
dynamical systems. Numerical continuation is a well-established
approach for comprehensive investigation of a model dynamics subject
to parameter variations \cite{KOG2007}, with particular focus on
the detection of parameter values for which the system behavior
suffers significant changes (bifurcations). In the present work, we
employ the continuation software COCO (Computational Continuation
Core \cite{DS2013}), a versatile development and analysis
platform oriented to the numerical treatment of continuation
problems solved via MATLAB. In particular, we will make extensive
use of the COCO-toolbox `hspo', which implements a set of numerical
routines for the path-following and bifurcation study of periodic
orbits of piecewise-smooth dynamical systems.

\begin{figure}[htbp!]
\centering
\psfrag{a}{\large(a)}\psfrag{b}{\large(b)}\psfrag{c}{\large(c)}\psfrag{J}{\Large$j$}\psfrag{V}{\Large$v$}
\psfrag{MJ}{\Large$j_{\mbox{\scriptsize
max}}$}\psfrag{B0}{\Large$\beta_{0}$}
\includegraphics[width=\textwidth]{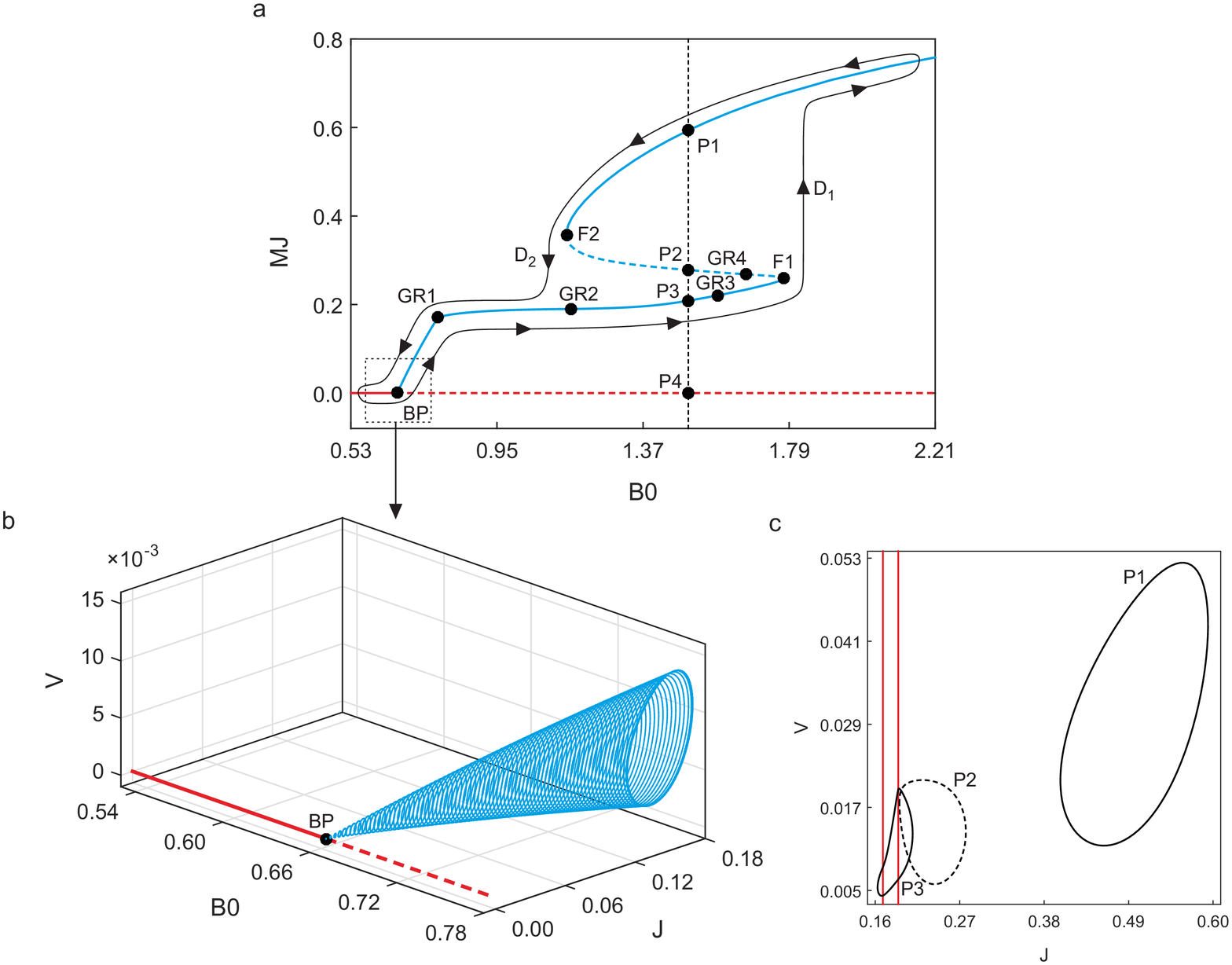}
\caption{(a) One-parameter continuation of the periodic response
shown in Fig.\ \ref{fig-bif-diag-zeta}(a) with respect to
$\beta_{0}$ (see \eqref{eq:newbeta}). In this diagram, the blue
curve represents continuation of periodic solutions, while the red
line stands for continuation of the disease-free equilibria. Solid
and dashed lines depict stable and unstable solutions, respectively.
The labels BP, F$i$ and GR$i$ represent a branching point, fold and
grazing bifurcations of limit cycles, respectively, while the labels
P$i$ correspond to coexisting solutions found for $\beta_{0}=1.5$
(see panel (c)). The closed curve D$_{1}$--D$_{2}$ represents
schematically a hysteresis loop of the system. Panel (b) shows the
solution manifold computed during the one-parameter continuation
around the bifurcation point BP.}\label{fig-bif-diag-B0}
\end{figure}

The point of departure for the numerical study is the periodic
solution shown in Fig.\ \ref{fig-bif-diag-zeta}(a) (inner diagram),
whereas previous sections solely afford the study on their
existence, not the stability. The periodic solutions are hereby
computed for the parameter values given in Tab.~\ref{tab:par}. Via
one-parameter continuation, we investigate how this solution is
affected by the amplitude of the seasonal forcing $\zeta$ (see
\eqref{seasonality}). The corresponding result is shown in Fig.\
\ref{fig-bif-diag-zeta}, which presents the behavior of the peak of
incidence level $j_{\mbox{\scriptsize max}}$ with respect to
$\zeta$. As analytically foreseen in Section~\ref{sec:approx}, the
peak of incidence level grows with $\zeta$. Note that larger $\zeta$
means higher seasonal variation of the vector population due to
meteorological factors (humidity, rainfall, temperature, etc.). As
explained in Section \ref{sec:pws}, the model \eqref{eq-pws} belongs
to the class of piecewise-smooth dynamical systems, which, in
contrast to smooth dynamical systems, can undergo the so-called
\emph{grazing} bifurcation \cite{BBC2004}. It occurs when a limit
cycle makes (quadratic) tangential contact with a discontinuity
boundary, defined by an event function introduced in Section
\ref{sec:pws}. In our case, there are two discontinuity boundaries,
given at $j=j^{*}$ and $j=c$. Consequently, during our investigation
we found three grazing bifurcations, located at
$\zeta\approx5.9192\times10^3$ (GR1), $\zeta\approx6.7175\times10^3$
(GR2) and $\zeta\approx8.0143\times10^3$ (GR3). Phase plots of the
corresponding grazing periodic solutions are depicted in Fig.\
\ref{fig-bif-diag-zeta}(b), where the tangential intersections with
the discontinuity boundaries can clearly be seen. A remarkable
feature of the bifurcation diagram shown in Fig.\
\ref{fig-bif-diag-zeta}(a) is the strong growth of the peak after
the grazing bifurcation GR2. This phenomenon is produced precisely
due to the crossing of the solution through the boundary $j=c$,
after which it is assumed that the impact of media vanishes or the
information has become tediously irrelevant for the susceptible
subpopulation, and therefore the force of infection increases
rapidly (see Fig.\ \ref{fig-beta-curve}).

Let us now investigate the behavior of the model \eqref{eq-pws} when
$\beta_{0}$ (see \eqref{eq:newbeta} and Fig.\ \ref{fig-beta-curve})
varies. The response curve can be seen in Fig.\
\ref{fig-bif-diag-B0}(a). To discuss the obtained result, let us
begin from the left part of the diagram, where $\beta_{0}$ is small.
If this is the case, the system presents an asymptotically stable
disease-free equilibrium represented by the solid red line. This
means that the incidence level among the population will decrease in
time until it disappears (for $t\to\infty$). During this regime, we
have that $\RM_0<1$ (see \eqref{eq:R0}). When $\RM_0$ becomes larger
than one (at $\beta_{0}\approx0.6473$), the disease-free equilibrium
loses stability, and a branch of endemic periodic solutions is born.
The latter gives rise to a branching point labeled BP (see Fig.\
\ref{fig-bif-diag-B0}(b)). Note that after this BP point, a
significant increment of the peak of incidence level can be
observed, until the grazing bifurcation GR1
($\beta_{0}\approx0.7795$) is found. This is then followed by a
sluggish increment of the endemic equilibrium state. The reason
behind is because after the bifurcation point GR1 (where the
periodic solution makes a tangential contact with the discontinuity
boundary $j=j^{*}$ from below), the solution crosses the threshold
$j=j^{*}$, at which it is assumed that the media give regular
reports about the disease spread among the susceptible population.
Therefore, people take preventive measures in such a way that the
infection rate decreases significantly (see Fig.\
\ref{fig-beta-curve}). If $\beta_{0}$ grows further, two additional
grazing points are found at $\beta_{0}\approx1.1624$ (GR2, grazing
contact with $j=c$) and $\beta_{0}\approx1.5837$ (GR3, grazing
contact with $j=j^{*}$ from above). For larger $\beta_{0}$, a
\emph{fold} bifurcation happening at the fold point F1
($\beta_{0}\approx1.7757$) is detected, where the periodic solution
becomes unstable. This unstable periodic response undergoes another
grazing bifurcation at $\beta_{0}\approx1.6660$ (GR4, grazing
contact with $j=c$) and recovers stability at the fold point F2
($\beta_{0}\approx1.1494$). After this point, the solution remains
stable with the increasing peak of incidence level as $\beta_{0}$
gets larger.

\begin{figure}[htbp!]
\centering \psfrag{a}{\large(a)}\psfrag{b}{\large(b)}
\psfrag{J}{\Large$j$}\psfrag{MJ}{\Large$j_{\mbox{\scriptsize
max}}$}\psfrag{jst}{\Large$j^{*}$}\psfrag{t}{\Large$t$}\psfrag{jm1}{\footnotesize$j_{\mbox{\tiny
max}}\approx0.59$}\psfrag{jm2}{\footnotesize$j_{\mbox{\tiny
max}}\approx0.21$}
\includegraphics[width=\textwidth]{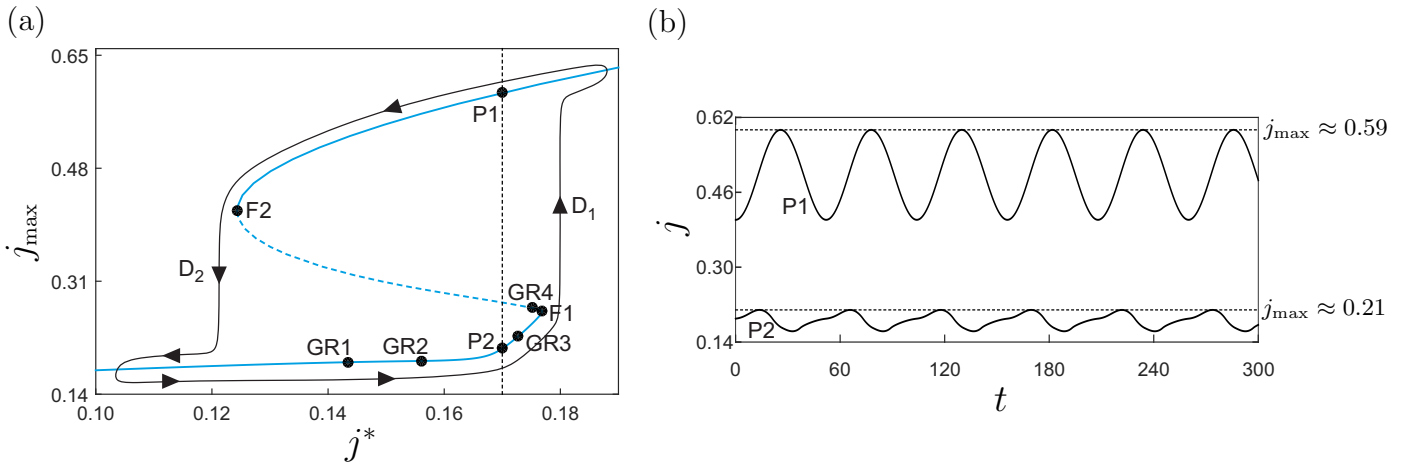}
\caption{(a) One-parameter continuation of the periodic response
shown in Fig.\ \ref{fig-bif-diag-zeta}(a) with respect to $j^{*}$
(see \eqref{eq:newbeta}). Labels and figure codes are defined as in
Fig.\ \ref{fig-bif-diag-B0}. (b) Time plot of two (coexisting)
stable periodic solutions computed for $j^{*}=0.17$ (P1,
P2).}\label{fig-bif-diag-jst}
\end{figure}

Another remarkable feature of the bifurcation diagram shown in Fig.\
\ref{fig-bif-diag-B0}(a) is the interplay between the fold
bifurcations F$1$ and F$2$ giving rise to \emph{hysteresis} in the
system, which schematically is represented by the closed curve
D$_{1}$--D$_{2}$ plotted in the figure. Moreover, the fold points
F$1$ and F$2$ define a parameter window for which the system
presents coexisting solutions (see Fig.\ \ref{fig-bif-diag-B0}(c)),
including the (unstable) disease-free equilibrium. A similar
scenario can be observed if now $j^{*}$ is considered as the
bifurcation parameter, see Fig.\ \ref{fig-bif-diag-jst}(a). As
before, a series of codimension-1 bifurcations are found along the
bifurcation diagram, located at $j^{*}\approx0.1434$ (GR1, grazing
contact with $j=j^{*}$ from above), $j^{*}\approx0.1561$ (GR2,
grazing contact with $j=c$), $j^{*}\approx0.1727$ (GR3, grazing
contact with $j=j^{*}$ from above), $j^{*}\approx0.1751$ (GR4,
grazing contact with $j=c$), $j^{*}\approx0.1769$ (F1, fold
bifurcation) and $j^{*}\approx0.1243$ (F2, fold bifurcation). As in
the previous case, the interaction between the fold points produces
a hysteresis loop, which in turn gives rise to the phenomenon of
\emph{multistability}, found for $j^{*}$ between F2 and F1. For
instance, at the initial value $j^{*}=0.17$, two stable (endemic)
periodic solutions can be found (at the test points P1 and P2),
depicted in Fig.\ \ref{fig-bif-diag-jst}(b). In the current
scenario, the solution at P2 can be identified as a ``desirable''
solution, owing to the low peak of incidence level
($j_{\mbox{\scriptsize max}}\approx0.21$). However, due to the
multistability, a sufficiently large perturbation to the system may
produce an undesired jump to the solution at P1, for which the peak
of incidence level is about three times larger
($j_{\mbox{\scriptsize max}}\approx0.59$), hence posing the risk of
a collapse of the available medical capabilities. In order to avoid
such an undesired scenario, one needs then to set $j^{*}$ as low as
possible (more precisely, below than $j^{*}\approx0.1243$ where the
fold bifurcation F2 occurs), which defines the point where the media
gives regular reports among the susceptible subpopulation thereby
encouraging people to take preventive measures.

\begin{figure}[H]
\centering
\psfrag{a}{\large(a)}\psfrag{b}{\large(b)}\psfrag{c}{\large(c)}\psfrag{d}{\large(d)}\psfrag{J}{\Large$j$}\psfrag{V}{\Large$v$}
\psfrag{logP}{\Large$\log\left(j_{\mbox{\scriptsize
p}}\right)$}\psfrag{logeps}{\Large$\log\left(\epsilon\right)$}\psfrag{slope}{\large
Slope $\approx$ -2}\psfrag{ec}{\large $\epsilon_{c}\approx0.0021$}
\includegraphics[width=\textwidth]{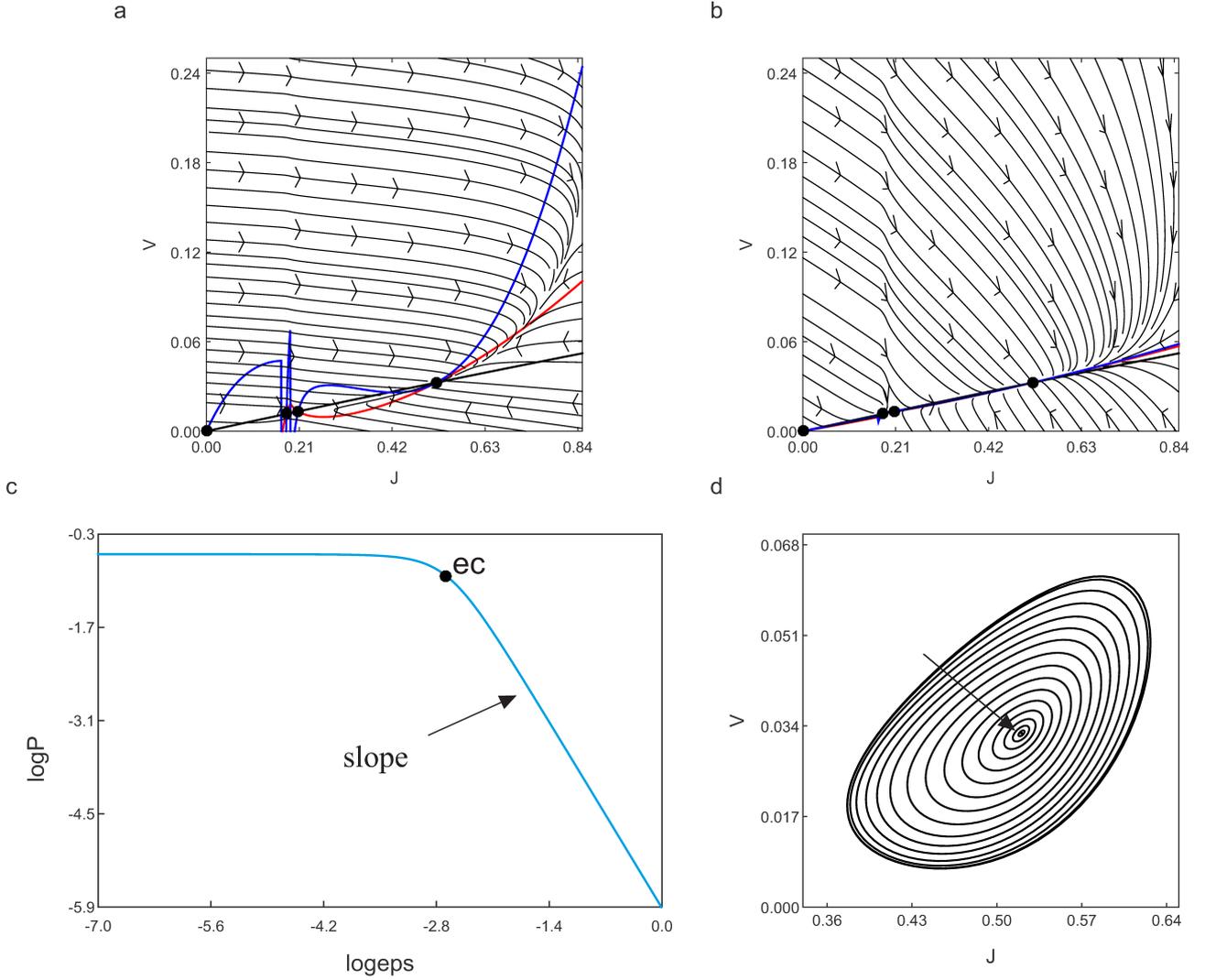}
\caption{Behavior of the model as the adiabatic parameter
(vector--host lifespan ratio) $\epsilon$ varies. Panels (a) and (b)
show phase portraits around the first-order (red curve),
second-order approximation of the slow manifold (blue curve) and the
critical manifold (black curve) computed for $\epsilon=10^{-2}$ and
$\epsilon=10^{-3}$, respectively. The small black circles encode the
existing equilibria, where the slow and critical manifold intersect.
Panel (c) shows the behavior of the peak-to-peak amplitude (see
\eqref{eq-jP}) of the periodic solution P1 displayed in Fig.\
\ref{fig-bif-diag-B0}(c). Panel (d) depicts the family of periodic
solutions obtained along the curve in (c). The arrow indicates the
direction of increasing $\epsilon$.}\label{fig-diag-epsilon}
\end{figure}

The last part of the numerical study consists of investigating the
behavior of the model as the adiabatic parameter $\epsilon$ varies.
The result is presented in Fig.\ \ref{fig-diag-epsilon} using
parameter values in Tab.~\ref{tab:par}. When the seasonal forcing is
deactivated, Section~\ref{sec:slowfast} has shown how nearby
solution trajectories approach the locally attractive slow manifold
($\epsilon>0$) before approaching the critical manifold and
ultimately stable equilibria, due to the slaving condition. Panels
(a) and (b) depict the preceding phenomenon for different
realizations of $\epsilon$. The first-order (red curve) and
second-order approximation of the slow manifold (blue curve) are
presented in the figure, whereby the two approximates and the actual
slow manifold meet and are close to the critical manifold as
$\epsilon$ is sufficiently small (Panel (b)). Panel (a) exclusively
shows how the higher-order terms surpass the influence of $\epsilon$
in the approximates, making them irregularly jumping around as
$\epsilon$ is rather large. This irregularity is worsened by the
fact that the model involves a piecewise-smooth vector field. Panels (c)
and (d) are obtained via one-parameter continuation of the periodic
solution P1 displayed in Fig.\ \ref{fig-bif-diag-B0}(c), using
$\epsilon$ as the control parameter. The numerical study reveals
that the (peak-to-peak) amplitude of the periodic solutions does not
vary significantly for $\epsilon$ small. This behavior, however,
notably changes when $\epsilon$ crosses a certain threshold
$\epsilon_{c}\approx0.0021$, when $\zeta=7560$. From this point on, the amplitude
decreases proportionally to $1/\epsilon^2$, as can be seen in Fig.\
\ref{fig-diag-epsilon}(c). This critical value can be computed from
\eqref{eq:M} as (neglecting transients and considering that
$\theta=\mu\slash \epsilon$ and $\rho=\rho_{\epsilon}\slash
\epsilon$), which is given by
$$\epsilon_{c}=10\frac{\mu\zeta}{\xi}.$$
This gives the $\epsilon$-value for which the constant component
($\bar{M}$) becomes ten times the amplitude of seasonality
($\zeta$). This means that for $\epsilon\geq\epsilon_{c}$, the
seasonal effects become less significant, and therefore the
amplitude of the periodic solutions decreases with $\epsilon$. The
quadratic decrement can be seen from \eqref{eq:theta}, \eqref{eq:h}
and \eqref{eq:jv}, hence producing a periodic excitation with
amplitude proportional to $1/\epsilon^2$. The preceding finding
provides additional novelty as apposed to the result in
Section~\ref{sec:approx} where $\epsilon\geq
\mu\zeta\slash\xi\sqrt{\theta^2+4\pi^2\omega^2}\approx 3.6\mu\zeta\slash\xi>\epsilon_c\slash 10$ makes the second-order
approximations eligible for replacing the actual periodic solutions.

\section{Conclusion}
In a rundown, the following work is done in the paper. We reduce and
rescale the simple host--vector, SISUV model into an IU model. A
periodic recruitment rate of the vector population is assigned to
take into account the influence of meteorological factors. The
argument about infected hosts kept in hospitals tells us that the
infection rate due to a contact between a susceptible vector and an
infected host is constant and sufficiently small. The other
infection rate from a typical contact between a susceptible host and
an infected vector is modelled as a function of the infected host
subpopulation. Under further assumptions, we designate a singular
perturbed system under the variation of the vector--host lifespan
ratio. In further examination, we center our work around
understanding the interplay between the media-triggered infection
rate, the periodic solutions surrounding equilibria of the
autonomous system, and the lifespan ratio variation. Two scenarios
for the infection rate are considered in this regard. The first
scenario puts forward a non-increasing infection rate, where the entire stability analysis returns a supercritical bifurcation. Moreover, a larger
alarming incidence level $j^{\ast}$ has been shown to give a larger
endemic equilibrium at the same magnitude of the basic reproductive
number. Keeping in mind the existence of the two notable equilibria,
two periodic solutions exist surrounding the equilibria, i.e.\ when
the seasonal forcing is activated. The second scenario presents a
new definition for the alarming incidence level $j^{\ast}$, i.e.\ a
maximal number of infected population share the available hospitals
can accommodate. This builds up an outline where the infection rate
decreases when $j>j^{\ast}$, but not long after at $c>j^{\ast}$,
poverty, reluctance in taking up preventive measures, despair,
tiresomeness, perceiving the disease as being easily curable, absence of medical access, and the exploding number of
hungrier vectors make the infection rate start to bounce up and gets
larger with the incidence level.

The following results are underlying regarding the second scenario.
We found a supercritical bifurcation when the basic reproductive
number is equal to one, also two fold bifurcations corresponding to
the switch from stable to unstable also from unstable to stable
endemic equilibrium branch. A similar result is also found from
numerical investigation over the stability of the existing periodic
solutions. For both autonomous and non-autonomous case, we acquire a
hysteresis loop. When the initial infection rate $\beta_0$ (or
eventually the basic reproductive number) increases, the presence of
\emph{overreaction} among the susceptible subpopulation attributed to
media reports helps suppress the endemicity level to a small order
of magnitude. Until then, $\beta_0$ is large enough that the
endemicity level jumps to a significantly larger value. We found that the closer $c$ and
$j^{\ast}$ are, the higher the possibility of encountering such a
hysteresis. In contrast, when $c$ is in the far right of $j^{\ast}$,
only a small endemic equilibrium can be obtained. All these mean
that overreaction is a bad response to the ever-increasing outbreak,
as a slow pace with sureness in the longevity and regularity in
applying preventive measures can have a better solution. A similar investigation over $j^{\ast}$ also gives a
hysteresis, therefore a sudden jump to a larger value in the endemicity level.
Notwithstanding the new definition for the alarming incidence level,
which the decision maker can always make up, we come to the same
conclusion as in the first model for the infection rate. We have
shown that overestimating alarming incidence level provides a good
solution to reduce the endemicity rather underestimating it, in any
way. When $j^{\ast}$ is sufficiently small, we have shown that no
jump to a blow-up is envisaged, also that the endemicity can be
suppressed as low as possible. As far as the lifespan ratio
$\epsilon$ is concerned, it turns to give us flexibility in
designing the model as to have solutions that may be comparable to
empirical data. For the sample case
$\mu\zeta\slash\xi\sqrt{\theta^2+4\pi^2\omega^2}\approx 3.6\mu\zeta\slash\xi\approx 7.56\times 10^{-4}<\epsilon=10^{-3}<\epsilon_c\approx 2.1\times 10^{-3}$,
we know that the critical manifold gives a quite good approximation to
the solution, whereby a second-order approximation for the periodic
solutions can be chosen for fast computation on an extremely large
time domain. We acquire both model reduction and small-order
approximation at the same time. Data assimilation using this model
with periodic datasets can be possible outlook.

\section*{Declaration of Competing Interest}
No conflict of interest exists in the submission of this manuscript.

\section*{Acknowledgements}
The second author has been supported by the DAAD Visiting
Professorships programme at the University of Koblenz-Landau. The third author is supported by the Ministry of Research, Technology and Higher Education of the Republic of Indonesia (Kemenristek DIKTI) with PUPT research grant scheme.

\bibliography{bibli}
\bibliographystyle{ieeetr}

\end{document}